\documentclass[11pt]{article}
\usepackage[dvips]{graphicx}
\begin{document}
\begin{center}
  {\Large {\bf Evolving Recursive Definitions with Applications to Dynamic Programming}}
\\[20pt] 
{\bf Keehang Kwon }\\
Dept. of Computer Engineering, DongA University \\
khkwon@dau.ac.kr
\end{center}

\newenvironment{describe}{\begin{list}{}{\setlength\leftmargin{80pt}}\setlength\labelsep{10pt}\setlength\labelwidth{70pt}}{\end{list}}

\newenvironment{flag}{\begin{list}{\makebox[20pt]{\hss$\circ$\enspace}}
                                  {\labelwidth 20pt}}{\end{list}}



\newenvironment{numberedlist}
{\begin{list}{\makebox[20pt]{\hss(\arabic{itemno})\enspace}}
             {\usecounter{itemno}\labelwidth 20pt}}{\end{list}}

\newenvironment{alphabetlist}
{\begin{list}{\makebox[20pt]{\hss(\alph{itemno1})\enspace}}
             {\usecounter{itemno1}\labelwidth 20pt}}{\end{list}}

\newenvironment{romanlist}
{\begin{list}{\makebox[20pt]{\hss(\roman{itemno2})\enspace}}
             {\usecounter{itemno2}\labelwidth 20pt}}{\end{list}}

\newcounter{itemno}

\newcounter{itemno1}

\newcounter{itemno2}
\newcounter{lemma}
\newcounter{exno}

\newcounter{defno}







\newenvironment{defn}{\refstepcounter{defno}\medskip \noindent {\bf
Definition \thedefno.\ }}{\medskip}

\newenvironment{ex}{\refstepcounter{exno}\medskip \noindent {\bf
Example \theexno.\ }}{\medskip}

\newcommand{\sep}{\;\vert\;}

\newcommand{\ra}{\rightarrow}
\newcommand{\app}{\ }
\newcommand{\appt}{\ }
\newcommand{\tup}[1]{\langle\nobreak#1\nobreak\rangle}

\newcommand{\hu}{{\cal H}^+}
\newcommand{\Free}{{\cal F}}
\newcommand{\oprove}{\vdash\kern-.6em\lower.7ex\hbox{$\scriptstyle O$}\,}
\newcommand{\true}{\top}

\newcommand{\Dscr}{{\cal D}}
\newcommand{\Pscr}{{\cal P}}
\newcommand{\Gscr}{{\cal G}}
\newcommand{\Fscr}{{\cal F}}
\newcommand{\Vscr}{{\cal V}}
\newcommand{\Uscr}{{\cal U}}
\newcommand{\pderivation}{{\cal P}\kern -.1em\hbox{\rm -derivation}}
\newcommand{\pderivationl}{{\cal P}\kern -.1em\hbox{\em -derivation}}
\newcommand{\pderivable}{{\cal P}\kern -.1em\hbox{\rm -derivable}}
\newcommand{\pderivablel}{{\cal P}\kern -.1em\hbox{\em -derivable}}
\newcommand{\pderivations}{{\cal P}\kern -.1em\hbox{\rm -derivations}}
\newcommand{\pderivability}{{\cal P}\kern -.1em\hbox{\rm -derivability}}
\newcommand{\eqm}[1]{=_{\scriptscriptstyle #1}}
\newcommand\subsl{\preceq}
\newcommand{\fnrestr}{\uparrow}

\newcommand{\match}{{\rm MATCH}}
\newcommand{\triv}{{\rm TRIV}}
\newcommand{\imit}{{\rm IMIT}}
\newcommand{\proj}{{\rm PROJ}}
\newcommand{\simpl}{{\rm SIMPL}}
\newcommand{\failed}{{\bf F}}

\newcommand{\Dsiginst}[1]{{[#1]_\Sigma}}
\newcommand{\Psiginst}[1]{{[#1]_\Sigma}}
\newcommand{\lnorm}{{\lambda}norm}
\newcommand{\dseq}[2]{#1_1,\ldots,#1_{#2}}

\newcommand{\all}{\forall}
\newcommand{\some}{\exists}
\newcommand{\lambdax}[1]{\lambda #1\,}
\newcommand{\somex}[1]{\some#1\,}
\newcommand\allx[1]{\all#1\,}

\newcommand{\subs}[3]{[#1/#2]#3}
\newcommand{\rep}[3]{S^{#2}_{#1}{#3}}
\newcommand{\ie}{{\em i.e.}}
\newcommand{\eg}{{\em e.g.}}

\newcommand{\lbotr}{$\bot$-R}
\newcommand{\ldbotr}{\bot\mbox{\rm -R}}
\newcommand{\landl}{$\land$-L}
\newcommand{\ldandl}{\land\mbox{\rm -L}}
\newcommand{\landr}{$\land$-R}
\newcommand{\ldandr}{\land\mbox{\rm -R}}
\newcommand{\lorl}{$\lor$-L}
\newcommand{\ldorl}{\lor\mbox{\rm -L}}
\newcommand{\lorr}{$\lor$-R}
\newcommand{\ldorr}{\lor\mbox{\rm -R}}
\newcommand{\limpl}{$\supset$-L}
\newcommand{\ldimpl}{\supset\mbox{\rm -L}}
\newcommand{\limpr}{$\supset$-R}
\newcommand{\ldimpr}{\supset\mbox{\rm -R}}
\newcommand{\lnegl}{$\neg$-L}
\newcommand{\ldnegl}{\neg\mbox{\rm -L}}
\newcommand{\ldnegr}{\neg\mbox{\rm -R}}
\newcommand{\lalll}{$\forall$-L}
\newcommand{\ldalll}{\forall\mbox{\rm -L}}
\newcommand{\lallr}{$\forall$-R}
\newcommand{\ldallr}{\forall\mbox{\rm -R}}
\newcommand{\lsomel}{$\exists$-L}
\newcommand{\ldsomel}{\exists\mbox{\rm -L}}
\newcommand{\lsomer}{$\exists$-R}
\newcommand{\ldsomer}{\exists\mbox{\rm -R}}
\newcommand{\ldlamlr}{\lambda}
\newcommand{\sequent}[2]{\hbox{{$#1\ \longrightarrow\ #2$}}}
\newcommand{\prog}[2]{\hbox{{$#1\ \supset\ #2$}}}
\newcommand{\run}{\Gamma}

\newcommand{\Ibf}{{\bf I}}
\newcommand{\Cbf}{{\bf C}} 
\newcommand{\Cbfpr}{{\bf C'}}

\newcommand{\cprove}{\vdash_C}

\newcommand{\iprove}{\vdash_I}

\newsavebox{\lpartfig}
\newsavebox{\rpartfig}


\newenvironment{exmple}{
 \begingroup \begin{tabbing} \hspace{2em}\= \hspace{3em}\= \hspace{3em}\=
\hspace{3em}\= \hspace{3em}\= \hspace{3em}\= \kill}{
 \end{tabbing}\endgroup}
\newenvironment{example2}{
 \begingroup \begin{tabbing} \hspace{8em}\= \hspace{2em}\= \hspace{2em}\=
\hspace{10em}\= \hspace{2em}\= \hspace{2em}\= \hspace{2em}\= \kill}{
 \end{tabbing}\endgroup}

\newcommand{\sand}{sand} 
\newcommand{\pand}{pand} 
\newcommand{\cor}{cor} 

\newcommand{\lb}{\langle}
\newcommand{\rb}{\rangle}
\newcommand{\pr}{prov}
\newcommand{\prG}{intp}
\newcommand{\prSG}{intp_E}
\newcommand{\intp}{intp_o}
\newcommand{\prove}{exec} 
\newcommand{\np}{invalid} 
\newcommand{\Ra}{\supset}  
\newcommand{\Cscr}{{\cal C}}
\newcommand{\seqweb}{SProlog}
\newcommand{\sprog}{{SProlog}}

\newcommand{\seqand}{\prec}
\newcommand{\seqor}{\cup}
\newcommand{\seqandq}[2]{\prec_{#1}^{#2}}
\newcommand{\parandq}[2]{\land_{#1}^{#2}}
\newcommand{\exq}[2]{\exists_{#1}^{#2}}
\newcommand{\ext}{intp_G}

\newcommand{\code}[1]{\ulcorner #1 \urcorner}
\newcommand{\mldi}{\hspace{2pt}\mbox{\footnotesize $\vee$}\hspace{2pt}}
\newcommand{\mlci}{\hspace{2pt}\mbox{\footnotesize $\wedge$}\hspace{2pt}}
\newcommand{\emptyrun}{\langle\rangle} 
\newcommand{\oo}{\bot}            
\newcommand{\pp}{\top}            
\newcommand{\xx}{\wp}               
\newcommand{\legal}[2]{\mbox{\bf Lr}^{#1}_{#2}} 
\newcommand{\win}[2]{\mbox{\bf Wn}^{#1}_{#2}} 
 \newcommand{\one}{\mbox{\sc One}}
 \newcommand{\two}{\mbox{\sc Two}}
 \newcommand{\three}{\mbox{\sc Three}}
 \newcommand{\four}{\mbox{\sc Four}}
 \newcommand{\first}{\mbox{\sc Derivation}}
 \newcommand{\second}{\mbox{\sc Second}}
 \newcommand{\uorigin}{\mbox{\sc Org}}
 \newcommand{\image}{\mbox{\sc Img}}
 \newcommand{\limitset}{\mbox{\sc Lim}}
 \newcommand{\fif}{\mbox{\bf CL15}}
\newcommand{\col}[1]{\mbox{$#1$:}}

\newcommand{\sti}{\mbox{\raisebox{-0.02cm}
{\scriptsize $\circ$}\hspace{-0.121cm}\raisebox{0.08cm}{\tiny $.$}\hspace{-0.079cm}\raisebox{0.10cm}
{\tiny $.$}\hspace{-0.079cm}\raisebox{0.12cm}{\tiny $.$}\hspace{-0.085cm}\raisebox{0.14cm}
{\tiny $.$}\hspace{-0.079cm}\raisebox{0.16cm}{\tiny $.$}\hspace{1pt}}}
\newcommand{\costi}{\mbox{\raisebox{0.08cm}
{\scriptsize $\circ$}\hspace{-0.121cm}\raisebox{-0.01cm}{\tiny $.$}\hspace{-0.079cm}\raisebox{0.01cm}
{\tiny $.$}\hspace{-0.079cm}\raisebox{0.03cm}{\tiny $.$}\hspace{-0.085cm}\raisebox{0.05cm}
{\tiny $.$}\hspace{-0.079cm}\raisebox{0.07cm}{\tiny $.$}\hspace{1pt}}}

\newcommand{\seq}[1]{\langle #1 \rangle}           


\newcommand{\mla}{\mbox{{\Large $\wedge$}}}
\newcommand{\mle}{\mbox{{\Large $\vee$}}}

\newcommand{\pst}{\mbox{\raisebox{-0.01cm}{\scriptsize $\wedge$}\hspace{-4pt}\raisebox{0.16cm}{\tiny $\mid$}\hspace{2pt}}}
\newcommand{\gneg}{\neg}                  
\newcommand{\mli}{\rightarrow}                     
\newcommand{\cla}{\mbox{\large $\forall$}}      
\newcommand{\cle}{\mbox{\large $\exists$}}        
\newcommand{\mld}{\vee}    
\newcommand{\mlc}{\wedge}  
\newcommand{\ade}{\mbox{\Large $\sqcup$}}      
\newcommand{\ada}{\mbox{\Large $\sqcap$}}      
\newcommand{\add}{\sqcup}                      
\newcommand{\adc}{\sqcap}                      

\newcommand{\tlg}{\bot}               
\newcommand{\twg}{\top}               
\newcommand{\st}{\mbox{\raisebox{-0.05cm}{$\circ$}\hspace{-0.13cm}\raisebox{0.16cm}{\tiny $\mid$}\hspace{2pt}}}
\newcommand{\cst}{{\mbox{\raisebox{-0.05cm}{$\circ$}\hspace{-0.13cm}\raisebox{0.16cm}{\tiny $\mid$}\hspace{1pt}}}^{\aleph_0}} 
\newcommand{\cost}{\mbox{\raisebox{0.12cm}{$\circ$}\hspace{-0.13cm}\raisebox{0.02cm}{\tiny $\mid$}\hspace{2pt}}}
\newcommand{\ccost}{{\mbox{\raisebox{0.12cm}{$\circ$}\hspace{-0.13cm}\raisebox{0.02cm}{\tiny $\mid$}\hspace{1pt}}}^{\aleph_0}} 
\newcommand{\pcost}{\mbox{\raisebox{0.12cm}{\scriptsize $\vee$}\hspace{-4pt}\raisebox{0.02cm}{\tiny $\mid$}\hspace{2pt}}}

\newcommand{\intimpl}{\mbox{\hspace{2pt}$\circ$\hspace{-0.14cm} \raisebox{-0.043cm}{\Large --}\hspace{2pt}}}
\newcommand{\sintimpl}{\mbox{\hspace{2pt}\raisebox{0.033cm}{\tiny $ | \hspace{-4pt} >$}\hspace{-0.14cm} \raisebox{-0.039cm}{\large --}\hspace{2pt}}}
\newcommand{\sst}{\mbox{\raisebox{-0.07cm}{\scriptsize $-$}\hspace{-0.2cm}$\pst$}}
\newcommand{\scost}{\mbox{\raisebox{0.20cm}{\scriptsize $-$}\hspace{-0.2cm}$\pcost$}}
\newcommand{\sqc}{\mbox{\hspace{2pt}\small \raisebox{0.0cm}{$\bigtriangleup$}\hspace{2pt}}}
\newcommand{\sqci}{\mbox{\scriptsize \raisebox{0.0cm}{$\bigtriangleup$}}}
\newcommand{\sqd}{\mbox{\hspace{2pt}\small \raisebox{0.06cm}{$\bigtriangledown$}\hspace{2pt}}}
\newcommand{\sqdi}{\mbox{\scriptsize \raisebox{0.05cm}{$\bigtriangledown$}}}
\newcommand{\sqe}{\mbox{\large \raisebox{0.07cm}{$\bigtriangledown$}}}
\newcommand{\sqa}{\mbox{\large \raisebox{0.0cm}{$\bigtriangleup$}}}
\newcommand{\tgd}{\mbox{\hspace{2pt}$\vee$\hspace{-1.29mm}\raisebox{0.1mm}{\rule{0.13mm}{2mm}}\hspace{5pt}}}    
\newcommand{\tgc}{\mbox{\hspace{2pt}$\wedge$\hspace{-1.29mm}\raisebox{0.02mm}{\rule{0.13mm}{2mm}}\hspace{5pt}}}    
\newcommand{\tge}{\hspace{1pt}\mbox{\Large $\vee$\hspace{-1.84mm}\raisebox{0.1mm}{\rule{0.13mm}{3.0mm}}\hspace{6pt}}}   
\newcommand{\tga}{\mbox{\hspace{1pt}\Large $\wedge$\hspace{-1.84mm}\raisebox{0.02mm}{\rule{0.13mm}{3.0mm}}\hspace{6pt}}}     
\newcommand{\tgpst}{\mbox{\raisebox{-0.01cm}{\scriptsize $\wedge$}\hspace{-4pt}\raisebox{0.06cm}{\small $\mid$}\hspace{2pt}}}
\newcommand{\tgpcost}{\mbox{\raisebox{0.12cm}{\scriptsize $\vee$}\hspace{-3.8pt}\raisebox{0.04cm}{\small $\mid$}\hspace{2pt}}}
\newcommand{\tgst}{\mbox{\raisebox{-0.05cm}{$\circ$}\hspace{-0.12cm}\raisebox{0.05cm}{\small $\mid$}\hspace{2pt}}} 
\newcommand{\tgcost}{\mbox{\raisebox{0.12cm}{$\circ$}\hspace{-0.12cm}\raisebox{0.04cm}{\small $\mid$}\hspace{2pt}}}


\newtheorem{theoremm}{Theorem}[section]
\newtheorem{conditionss}{Condition}[section]
\newtheorem{thesiss}[theoremm]{Thesis}
\newtheorem{definitionn}[theoremm]{Definition}
\newtheorem{lemmaa}[theoremm]{Lemma}
\newtheorem{notationn}[theoremm]{Notation}
\newtheorem{propositionn}[theoremm]{Proposition}
\newtheorem{conventionn}[theoremm]{Convention}
\newtheorem{examplee}[theoremm]{Example}
\newtheorem{remarkk}[theoremm]{Remark}
\newtheorem{factt}[theoremm]{Fact}
\newtheorem{exercisee}[theoremm]{Exercise}
\newtheorem{questionn}[theoremm]{Open Problem}
\newtheorem{conjecturee}[theoremm]{Conjecture}

\newenvironment{exercise}{\begin{exercisee} \em}{ \end{exercisee}}
\newenvironment{definition}{\begin{definitionn} \em}{ \end{definitionn}}
\newenvironment{theorem}{\begin{theoremm}}{\end{theoremm}}
\newenvironment{lemma}{\begin{lemmaa}}{\end{lemmaa}}
\newenvironment{proposition}{\begin{propositionn} }{\end{propositionn}}
\newenvironment{convention}{\begin{conventionn} \em}{\end{conventionn}}
\newenvironment{remark}{\begin{remarkk} \em}{\end{remarkk}}
\newenvironment{proof}{ {\bf Proof.} }{\  \rule{2.5mm}{2.5mm} \vspace{.2in} }
\newenvironment{idea}{ {\bf Idea.} }{\  \rule{1.5mm}{1.5mm} \vspace{.15in} }
\newenvironment{example}{\begin{examplee} \em}{\end{examplee}}
\newenvironment{fact}{\begin{factt}}{\end{factt}}
\newenvironment{notation}{\begin{notationn} \em}{\end{notationn}}
\newenvironment{conditions}{\begin{conditionss} \em}{\end{conditionss}}
\newenvironment{question}{\begin{questionn}}{\end{questionn}}
\newenvironment{conjecture}{\begin{conjecturee}}{\end{conjecturee}}

\newcommand{\muprolog}{{$OOP^n$}}

  Inspired by computability logic\cite{Jap03}, we  refine recursive function definitions into two kinds:
blindly-quantified (BQ) ones and parallel universally quantified (PUQ) ones. BQ definitions
corresponds to the
traditional ones where recursive definitions are $not$ evolving.
PUQ definitions are {\it evolving} in the course of computation, leading to
 automatic memoization.
In addition, based on this idea, we propose a new, high-level
 object-oriented language \muprolog.   The merit of 
 this language is that it is a simplest object-oriented programming.

keywords:
 evolving recursive functions, dynamic programming.

\renewcommand{\intp}{eval} 


\section{Introduction}\label{sec:intro}

The theory of recursive functions  provides a basis for popular
functional/imperative languages such as ML, Java and Javascript.
It includes operations of composition, recursion, $etc$. Although the
theory
is quite expressive,
it does not support automatic memoization and dynamic programming
in a high-level way.
As a consequence, dynamic programming is known to be cumbersome, as
it requires the programmer to write
extra, messy code to deal with memoization.
In other words, it is a low-level
approach.

In this paper, we propose a high-level approach to this problem.
The idea is based on extending recursive definitions
with some evolutionary features, which we call {\it evolving}
recursive definitions.
This idea comes from the recent important work called
Computability Logic(CoL). See for its details\cite{Jap03}.

\section{Two Interpretations of $\cla x$}\label{sec:intro}

There are two different ways to interpret

\[ \cla xf(x) = E(x) \]
\noindent
  where $E$ is an expressions.

  \begin{itemize}

  \item The $parallel$ interpretation is the following:
    the above is true if it is true for all terms, i.e.,

    \[ f(0)=E(0)\mlc f(1) = E(1) \mlc \ldots \]

  \item The $blind$ interpretation is the following:
    the above is true if it is true independent of $x$.
    This is a subtle concept and can be viewed as an $internally$ parallel interpretation. Note that this
    definition  cannot be
     expanded.
  \end{itemize}
  
  \noindent
  In the sequel, we use $\mla x$ instead of $\cla x$
  to denote parallel universal quantifier.
  Now the difference is the following:
  
Evaluating  $f(c)$ with respect to  $\cla x f(x) = E(x)$
has the following intended semantics:
create $f(c) = E(c)$, evaluate $E(c)$ and then discard $f(c) = E(c)$.

In contrast, evaluating  $f(c)$ with respect to  $\mla x f(x) = E(x)$
has the following intended semantics:
create $f(c) = E(c)$, evaluate $E(c)$ and then do not
discard $f(c) = E(c)$.

Thus, $\mla x$ definitions are evolving and intended to perform automatic memoization.
This is a novel feature which is not present in
traditional recursive definitions.
On the other hand,  the blind
interpretation is the one that is used in traditional recursive definition.

\section{Fibonacci with Memoization}

In the traditional  approach, function evaluation is based
on the blind quantification. 
For example, a  $fib$ function is  specified as: \\

$   fib(0) = 1 $

$   fib(1) = 1$

$   \forall x (fib(x\!+\!2) = fib(x\!+\!1)+f(x)). $ \\

Note that the third definition is a  {\it blind-quantified
  definition}  in \cite{Jap03} -- means that it
is true independent of $x$.

Now, to compute $fib(3)$, our machine {\it temporarily}
creates the following new instances in the course of evaluation and discards them after.

$   fib(3) = f(3)+f(2) = 3$

$   fib(2) = f(1)+f(0) = 2. $

To support (top-down) automatic memoization, all we have to do is to replace $\cla x$
by $\mla x$.

In the sense of \cite{Jap03}, PUQ definitions require to
 generates new instances of the existing function
definitions and adds them permanently in front of the program. These new instances
play a role similar to automatic memoization.

For example, a  $fib$ function can be specified as: \\

$   fib(0) = 1 $

$   fib(1) = 1$

$   \mla x (fib(x\!+\!2) = fib(x\!+\!1)+f(x)). $ \\

Note that the third definition  is compressed
and  needs to be expanded during execution. 
For example, to compute $fib(3)$,  the above definition evolves/expands  to the following \\

$      fib(2) = 2 $

$ fib(3) = 3$ 

$   fib(0) = 1 $

$   fib(1) = 1$

$   \mla x (fib(x\!+\!2) = fib(x\!+\!1)+f(x)). $ \\

One consequence of our approach is that
it supports dynamic programming.

\section{The Language}\label{s:logic}

The language is a version of the core functional languages --- also one of recursive functions ---
 with PUQ expressions. 
It is described
by $E$- and $D$-rules given by the abstract syntax as follows:
\begin{exmple}
\>$E ::=$ \>  $c \sep x \sep  h(E,\ldots,E) \sep     \top $ \\   \\
\>$D ::=$ \>  $\cla \vec{x} f(\vec{x}) = E   \sep \mla \vec{x} f(\vec{x}) = E   \sep  D \land D  $\\
\end{exmple}
\noindent
In the abstract syntax, $E$ and $D$ denote the expressions and the definitions, respectively.
In the rules above, $c$ is a constant, $x$ is a variable, $t$ is a term which is either a variable or a constant.
  $D$ is called a program in this language.

\newcommand{\bc}{bc}

Following the traditional approach for defining semantics \cite{Mil89jlp}, we will present the semantics of this language, essentially an interpreter for the language, as a set of rules in Definition 1.
The evaluation strategy assumed by these rules is an eager evaluation. 
 Note that execution  alternates between 
two phases: the evaluation phase defined by \textit{eval}
and the backchaining phase by \textit{bc}. 

In  the evaluation phase, denoted by $\intp(D,E,K,D')$, the machine tries to evaluate an expression $E$ from the program $D$ to get a value $K$ and an evolved definition $D'$. The rules (7) -- (10) 
are related to this phase.
For instance, if $E$ is a function call $h$, the machine first evaluates all of its arguments and then looks for a definition of $h$ in the program in the backchaining mode (Rule 6).

The rules (1) -- (6) describe the backchaining mode, denoted by $bc(D_1,D,h,K,D')$.
In the backchaining mode, the machine tries 
to evaluate a function call $h$
by using the function definition in the program $D_1$. $K$ is the value after evaluation and
$D'$ is what $D$ evolves to.  $bc_b$ takes care of the BQ definitions and $bc_p$ takes care of the PUQ
ones.

\begin{defn}\label{def:semantics}
Let $E$ be an expression  and let $D$ be a program.
Then the notion of   evaluating $\lb D,E\rb$ to a value $K$ and getting a new definition $D'$ --- $\intp(D,E,K,D')$ --- 
 is defined as follows:

\begin{numberedlist}


\item    $\bc(h(c_1,\ldots,c_n) = E, D, h(c_1,\ldots,c_n), K, D')$ \\ if 
  $\intp(D, E, K, D')$ \% switch to evaluation mode.

\item    $\bc_b(h(c_1,\ldots,c_n) = E, D, h(c_1,\ldots,c_n), K,
  D')$ \\ if  $\intp(D, E, K, D') $ \% memoization.
  
  \item    $\bc_p(h(c_1,\ldots,c_n) = E, D, h(c_1,\ldots,c_n), K, 
  (h(c_1,\ldots,c_n) = K)::D')$ \\ if  $\intp(D, E, K, D') $ \% memoization.

\item    $\bc(D_1\land D_2,D,h(c_1,\ldots,c_n),K)$  \\
 if   $\bc(D_i,D,h(c_1,\ldots,c_n),K)$. 
 provided that $D_i(i = 1,2)$ is the first matching declaration.

\item    $\bc(\cla x_1,\ldots,\cla x_n h(x_1,\ldots,x_n) = E, D, h(c_1,\ldots,c_n),K,D')$ \\
if   $\bc_b(h(c_1,\ldots,c_n) = E(c_1,\ldots,c_n), D, h(c_1,\ldots,c_n),K,D')$ . \% argument passing to
 $h$ and $E$. \% blind quantification

\item    $\bc(\mla x_1,\ldots,\mla x_n h(x_1,\ldots,x_n) = E, D, h(c_1,\ldots,c_n),K,D')$ \\
if   $\bc_p(h(c_1,\ldots,c_n) = E(c_1,\ldots,c_n), D, h(c_1,\ldots,c_n),K,D')$ \% argument passing to
 $h$ and $E$. \% parallel universal quantification

\item    $\intp(D,h(c_1,\ldots,c_n),K)$ \\ if   $\bc(D,D,h(c_1,\ldots,c_n),K )$. \% 
 switch to backchaining by making a copy of $D$ for a function call.

\item    $\intp(D,h(E_1,\ldots,E_n),K,D')$ \\ if $\intp(D,E_i,c_i,D_i)$ and 
$\intp(D_1\mlc\ldots \mlc D_n,h(c_1,\ldots,c_n),K,D')$.
 \%  evaluate the arguments first.

\item    $\intp(D,\top,\top,D)$. \% 
 $\top$ is always a success.

\item   $\intp(D, c, c, D)$.  \% A success if $c$ is a constant.


\end{numberedlist}
\end{defn}

\section{Object-Oriented Programming}

Evolving definitions in the previous section perform sequential search for a function definition.
This is a slow process! We can  speed up this search by permitting  {\it locations}
for a set of function definitions.  The addition of locations makes it possible to 
provide a direct lookup of a function.

For example, assume that
$fib(X) = E$ is stored at a location $/a[X]$.  Then $fib$ can be
rewritten as: \\

 $/a[1]: fib(1) =1$

 $/a[2]: fib(2) = 1$

 $\mla x\ /a[x\!+\!2]: fib(x\!+\!2) = /a[x\!+\!1].fib(x\!+\!1)+ /a[x].fib(x)$

 $\cla n\ /fib: fib(n) = /a[n].fib[n].$ \\

Now consider an expression $/fib.fib(4)$. The
machine creates two instances at run time. \\

$ /a[3].fib(3) = 2$

$  /a[4].fib(4) = 3$ \\

Note that the above is nothing but an object-oriented
programming in a distilled form. That is, in object-oriented terms, 
$/a[1],/a[2]$ are regular objects and  $\mla x\ /a[x\!+\!2]$ is a {\it class}
object.  Our language has
some interesting features:

\begin{enumerate}
  
\item Instances are created lazily and automatically.
  
\item As we will see later, it supports nested objects.
  
\item Class is not a primary means for creating an object: objects can be created without introducing
  its class.

\end{enumerate}

Finally, note that adding imperative features to our languages poses no problem, still leading to a very concise code.

\section{Nested objects}

Consider an object definition of the form

\begin{exmple}
$/a:$ \> $g(1) = E $    \\
 \> $g(X\!+\!1) = E_1 $     \\
\end{exmple} 

\noindent For speedup, the method $g$ within an object $/a$  can be further refined to

 \begin{exmple}
$/a:$ \> $/b[1]: g(1) = E $    \\
 \> $/b[X\!+\!1]: g(X\!+\!1) = E_1 $    
 \end{exmple}

 \noindent Now each call to $g$ must be of the form $/a./b[].g$.
Note that  our language supports a novel
 {\it nested} object-oriented programming.
Nested objects are quite useful for further 
speedup as well as clustering objects.
We will look into this in the future.

Finally, our language is clearly influenced. by computability
logic web\cite{Jap03,Jap08} (CoLweb) which is a promising approach to
reaching general AI.  Ideas in this paper -- class agents and nested agents  -- will be useful for organizing agents and
their knowledgebases in CoLweb.

\bibliographystyle{ieicetr}

\end{document}